\documentstyle[twocolumn,aps,epsfig]{revtex}

\newcommand{\bq}{\begin{equation}}
\newcommand{\eq}{\end{equation}}
\newcommand{\bqa}{\begin{eqnarray}}
\newcommand{\eqa}{\end{eqnarray}}
\newcommand{\nn}{\nonumber \\}

\begin{document}
\draft
\title{On the Origin of Peak-dip-hump Structure in the  
In-plane Optical Conductivity of the High $T_C$ Cuprates;
Role of Antiferromagnetic Spin Fluctuations of Short Range Order}
\author{Sung-Sik Lee, Jae-Hyeon Eom, 
Ki-Seok Kim and Sung-Ho Suck Salk$^{1,2}$}
\address{$^1$ Department of Physics,
Pohang University of Science and Technology,
Pohang, Kyoungbuk, Korea 790-784}
\address{$^2$ Korea Institute of Advanced Study, Seoul, 130-012 Korea}
\date{\today}

\maketitle

\begin{abstract}
An improved U(1) slave-boson approach 
is applied to study
the optical conductivity of the two dimensional
systems of antiferromagnetically correlated electrons
over a wide  range of hole doping and temperature.
Interplay between the spin and charge degrees of freedom is
discussed to explain the origin of the peak-dip-hump
structure in the in-plane conductivity of high $T_C$
cuprates.
The role of 
spin fluctuations  of short range
order(spin singlet pair) is investigated.
It is shown that 
the spin fluctuations 
of the short range order
can cause the mid-infrared hump,
 by exhibiting a linear increase of the
hump frequency with the antiferromagnetic Heisenberg coupling strength.
\\
\end{abstract}



High $T_C$ superconductors are the systems of
strongly correlated
electrons which show two dimensionality in charge transport.
Various levels of gauge theoretic
slave-boson approach to t-J Hamiltonian have been
proposed to study high $T_C$ 
superconductivity\cite{KOTLIAR,FUKUYAMA,UBBENS92,WEN,LEE,SSLEE,GIMM}.
Recently we proposed an SU(2)  
slave-boson theory\cite{SSLEE} which incorporated coupling
between the charge and spin degrees of freedom
into the Heisenberg term.
  The predicted phase diagram showed an arch-shaped
bose condensation line
in agreement with observation.
Using an improved U(1) slave-boson theory over
our earlier one\cite{GIMM},
 in this paper we study 
the cause of  peak-dip-hump structures 
of observed optical conductivity\cite{ROTTER,ROMERO,UCHIDA,PUCHKOV,LIU}. 
Various theories 
have been proposed to explain 
the cause of the peak-dip-hump
structure in the optical conductivity\cite{STOJKOVIC,MUNZAR,HASLINER}.
However, most studies  have been made  to a 
limited  range of hole doping and temperature,
based on empirical parameters deduced from measurements such as 
the inelastic neutron scattering(INS) and
the angle resolved photoemission spectroscopy(ARPES) 
data.

Using the nearly antiferromagnetic   
Fermi-liquid theory, 
Stojkovi$\acute{c}$ and Pines\cite{STOJKOVIC} reported a
study of normal state
optical conductivity for optimally doped and overdoped systems.
They showed
that the highly anisotropic 
scattering rate in different regions of
the Brillouin zone 
 leads to an 
average relaxation rate of 
the marginal Fermi-liquid  form. 
Their computed 
optical conductivity 
agreed well with experimantal data for the normal state of an 
optimally doped sample. 
Using the spin-fermion model\cite{MBP,CM} and
spin susceptibility parameters obtained from
INS and NMR,
Munzar, Bernhard and  Cardona\cite{MUNZAR} calculated the 
in-plane optical conductivity 
of optimally doped YBCO. 
Their study showed a good agreement with the observed
 peak-dip-hump structure 
at optimal doping. 
From the computed self energy
they showed that the hump is originated from
 the hot quasiparticles and the Drude peak,
 from the cold quasiparticles.
Haslinger, Chubukov and Abanov\cite{HASLINER} reported 
 optical conductivities $\sigma (\omega)$ of optimally doped cuprates
in the normal state 
by allowing coupling 
between the spin-fermions and the bosonic spin fluctuations. 
They found that the width of the peak in spectral function
$A_{\bf k}(\omega)$ scales linearly
with $\omega$ in both hot and cold spots in the Brillouin zone 
and $\sigma (\omega)$ is inversely linear in $\omega$ up to very high frequencies.

Various  studies 
 have been limited to a restricted range of hole doping 
and temperature,
 relying on empirical parameters deduced from INS and ARPES.
It is thus of great interest to resort to a 
theory which depends least on  empirical parameters 
and fits
for a wide  range both of hole doping(including the 
important region of underdoping)  and 
temperature(encompassing the pseudogap phase and the
superconducting phase).
For this cause we use an improved slave-boson theory of the 
t-J Hamiltonian\cite{SSLEE}
that we developed recently.

Here we briefly discuss the slave-boson theory
to discuss the coupling between the
spin and charge degrees of freedom\cite{SSLEE}.
The t-J Hamiltonian in the presence of the external 
electromagnetic field ${\bf A}$
is written,
\vspace{-0.08cm}
\bqa
H &=& -t \sum_{<i,j>} ( e^{i A_{ij}} \tilde{c}^{\dagger}_{i \sigma} 
\tilde{c}_{j \sigma} + H.C. ) \nn
&&+J \sum_{<i,j>}( {\bf S}_i \cdot {\bf S}_j - 
\frac{1}{4} {n}_{i} {n}_{j}) 
- \mu \sum_{i, \sigma}c^\dagger_{i \sigma} c_{i \sigma},
\label{t-J_Hamiltonian}
\eqa 
\vspace{-0.08cm}
with ${\bf S}_i = 
\frac{1}{2} \sum_{\alpha \beta }c_{i\alpha}^{\dagger} {\sigma}_{\alpha \beta} c_{i\beta} $.
Here $ A_{ij}$ is the external electromagnetic vector
potential;
$\tilde{c}_{i \sigma}( \tilde{c}^{\dagger}_{i \sigma})$,
the electron annihilation(creation) operator 
at each site and 
$ {\sigma}_{\alpha \beta} $, the Pauli spin matrix.
Rewriting the electron operator as a composite of 
spinon($f$) and holon($b$) operators, $c_{i \sigma} = f_{i \sigma} b^{\dagger}_i$
with the single occupancy constraint, 
$b^{\dagger}_i b_i + \sum_{\sigma} f^{\dagger}_{i \sigma}f_{i \sigma}  =1$,
we obtain the partition function, 
\vspace{-0.08cm}
\bqa
Z = \int 
{\cal D}f 
{\cal D}b
{\cal D}\lambda
 e^{-\int_{0}^{\beta} d \tau {\cal L}},
\eqa
\vspace{-0.08cm}
with ${\cal L}= \sum_{i, \sigma } f^{*}_{i \sigma}
  \partial_{\tau} f_{i \sigma} 
+  \sum_i b^{*}_{i} \partial_{\tau} b_{i}  + H_{t-J}$ 
where $H_{t-J}$ is the U(1) slave-boson representation of the above t-J
Hamiltonian(Eq.(\ref{t-J_Hamiltonian})), 
\vspace{-0.08cm}
\bqa
H_{t-J} & = & -t\sum_{<i,j>}(e^{i A_{ij}}
f_{i\sigma}^{\dagger}f_{j\sigma}b_{j}^{\dagger}b_{i} 
+ c.c.) \nonumber \\
&& -\frac{J}{2} \sum_{<i,j>} b_i b_j b_j^{\dagger}b_i^{\dagger}
(f_{i\downarrow}^{\dagger}f_{j\uparrow}^{\dagger}-f_{i\uparrow}^ {\dagger}
f_{j\downarrow}^{\dagger})(f_{j\uparrow}f_{i\downarrow}-f_{j\downarrow}
f_{i\uparrow}) \nn
&& - \mu\sum_{i,\sigma}  f_{i\sigma}^{\dagger}f_{i\sigma}
+ i\sum_{i} \lambda_{i}(f_{i\sigma}^{\dagger}f_{i\sigma}+b_{i}^{\dagger}b_{i} -1).
\label{Hamiltonian_fb}
\eqa
\vspace{-0.08cm}
This Hamiltonian can reaily be derived from the SU(2) theory\cite{SSLEE}.

From the Hubbard-Stratonovich
transformations involving hopping, spinon pairing and holon pairing
orders we obtain
 the partition function, 
\vspace{-0.08cm}
\bqa
Z = \int
{\cal D}f 
{\cal D}b
{\cal D}\chi
{\cal D}\Delta^f
{\cal D}\Delta^b
{\cal D}\lambda
e^{-\int_{0}^{\beta} d \tau {\cal L}_{eff}},
\label{free_energy_final}
\eqa
\vspace{-0.08cm}
with ${\cal L}_{eff} = {\cal L}^f + {\cal L}^b + {\cal L}_0$ is
the Lagrangian where 
${\cal L}^f  
= \sum_{i,\sigma} f^\dagger_{i, \sigma} \partial_{\tau} f_{i, \sigma}
-\frac{J(1- \delta)^2}{4}\sum_{<i,j>}
\Big\{ \chi_{ij} f^\dagger_{i, \sigma} f_{j, \sigma} + H.C. \Big\}
- \frac{J(1- \delta)^2}{2} \sum_{<i,j>}
\Big\{ \Delta^f_{ij} ( f_{i , \downarrow} f_{j , \uparrow}
- f_{i , \uparrow} f_{j , \downarrow}) 
+ H.C. \Big\}  $ for the spinon sector, 
$ {\cal L}^b = 
\sum_{i} b^\dagger_{i} \partial_{\tau} b_{i}
 -t \sum_{<i,j>}
\Big\{
e^{i A_{ij}} \chi_{ij} b^\dagger_{i} b_{j} + H.C.
\Big\}
-\frac{J}{2} \sum_{<i,j>}
|\Delta^f_{ij}|^2 \Big\{
\Delta^b_{ij} b^\dagger_i b^\dagger_j + H.C.
\Big\}$ for the holon sector and
${\cal L}^0 = J(1- \delta)^2 \sum_{<i,j>} \Big\{ 
|\Delta^f_{ij}|^2 + 
\frac{1}{4} |\chi_{ij}|^2 + \frac{1}{4}n_i \Big\}
+ \frac{J}{2} \sum_{<i,j>}
|\Delta^f_{ij}|^2 |\Delta^b_{ij}|^2 $. 
Here  $\chi$, ${ \Delta }^f$ and ${ \Delta }^b$ are
the hopping, spinon pairing and holon pairing order parameters
respectivity.

We obtain the optical conductivity $\sigma (\omega)$
and 
 the current response function $\Pi (\omega)$ of
 an isotropic 2-D mediun
in the external electric field ${\bf E}( \omega )$
by evaluating 
the second derivative of the free energy with respect to the
external vector potential $\bf{A}$,
\vspace{-0.08cm}
\small{
\bqa
\sigma ( \omega ) 
= \left. \frac{ \partial J_x (\omega) }{ \partial E_x (\omega)} \right|_{ E_x =0} 
=- \frac{1}{ i \omega } 
\left. \frac{ \partial^2 F  }{ \partial {A_x}^2} 
\right|_{A_x =0} 
= \frac{ \Pi_{xx} (\omega) }{ i \omega},
\eqa
}
\normalsize
\vspace{-0.08cm}
where $J_x$ is the induced current in the x direction;
 $F = - k_B T \ln Z $, the free energy  and
$\Pi_{xx} = - \left. \frac{ \partial^2  F }{ \partial {A_x}^2} \right|_{A_x =0}$,
the current response function in the x-direction. 
The total response function,
$\Pi=\Pi^{P}+\Pi^{D}$ is the sum of the
paramagnetic response function given by the
current-current correlation function $\Pi^{P}(r^\prime -r, t^\prime -t)
= <j_x(r^\prime ,t^\prime)
j_x(r,t )>-<j_x(r^\prime ,t^\prime)><j_x(r,t )>$ 
with the current operator $j_x(r,t) = it(c^\dagger_{r+x, \sigma}(t) c_{r,\sigma}(t )
- c^\dagger_{r,\sigma}(t) c_{r+x, \sigma}(t )   )$
and the diamagnetic response function
associated with the average kinetic energy,
$\Pi^{D}=<K_{xx}>= \left< -t \sum_{i,\sigma}
(c^\dagger_{i, \sigma} c_{i+x, \sigma} + H.C.) \right> $\cite{DAGOTTO}.

The phase difference per unit lattice spacing associated with the 
 hopping order parameter 
$\chi_{ij} = |\chi_{ij}|e^{a_{ij}}$ 
defines the gauge field,
 $a_{ij}= \partial_{ij}\theta = \theta_i - \theta_j$.
 The gauge fluctuations 
allow the back-flow condition 
in association with an interplay between the
charge and spin degrees of freedom 
orriginated from the effective 
 kinetic energy term of the t-J Hamiltonian. 
The effects of 
spin degrees of freedom 
are manifested  through 
the antiferromagnetic
spin fluctuations 
which appear in the Heisenberg exchange coupling term.
The antiferromagnetic
spin fluctuations of short range
order(spin singlet pair) occur through the presence of
correlations between adjacent electron spins.
We consider both the 
amplitude fluctuations of the
spinon pairing(spin singlet) order parameter $|{ \Delta }^f|$
and the gauge field fluctuations.
We first integrate out the spinon and holon fields
and take  the
saddle point value with respect to the holon pairing order 
parameter, spinon pairing order parameter phase,
the amplitude of hopping order parameter and
the Lagrangian multiplier fields in 
Eq.(\ref{free_energy_final}). We then obtain,
\small{
\vspace{-0.08cm}
\bqa
F[{\bf A}] 
        &=& - k_B T \ln
        \int  
        {\cal D} f {\cal D} b
         {\cal D} \chi
        {\cal D} \Delta^f
        {\cal D} \Delta^b
        {\cal D} \lambda
        e^{-\int_{0}^{\beta} d \tau 
        \big(
        {\cal L}^f + {\cal L}^b + {\cal L}_0 
        \big) } \nn
        & \approx & - k_B T \ln 
        \int {\cal D} {\bf a} {\cal D} | { \Delta }^f |
        \Big( e^{- \beta \big (F^f[{\bf a}, | { \Delta }^f |] 
        + F^b[{\bf A}, {\bf a}, | { \Delta }^f |]} \nn
        && {~~~~~~~~~~~~~~~~~~~~~~~~~~~~~ }
 \times {e}^{ F_0[{\bf a}, | { \Delta }^f |] \big)} \Big),
\eqa
}\normalsize
\vspace{-0.08cm}
where $ F^f = - k_B T \ln \int {\cal D} f 
e^{-\int_{0}^{\beta} d \tau {\cal L}^f} $ is
the spinon free energy,
 $ F^b = - k_B T \ln \int {\cal D} b
e^{-\int_{0}^{\beta} d \tau {\cal L}^b} $,
the holon free energy and
$F_0 = - k_B T \ln
e^{-\int_{0}^{\beta} d \tau {\cal L}_0} $. 
The external electromagnetic field
 couples only to the holon field 
but not to the spinon field.

Considering the gauge and antiferromagnetic
spin fluctuations up to second order 
we obtain the current response function,
\vspace{-0.08cm}
\bqa
\Pi &=& 
         \frac{\Pi^f \Pi^b}{\Pi^f + \Pi^b} 
+ \frac{ 
\left( \Pi^b_{a \Delta} - 
       \frac{\Pi^b_{a \Delta} + \Pi^f_{a \Delta}}
       {\Pi^b + \Pi^f} \Pi^b    
\right)^2 }
{   2  \frac{   ( \Pi^b_{a \Delta } + \Pi^f_{a \Delta } )^2}
      { \Pi^b + \Pi^f }  
- ( \Pi^0_{\Delta \Delta} + \Pi^b_{\Delta \Delta} + \Pi^f_{\Delta \Delta} )   }, 
\label{Pi}
\eqa
\vspace{-0.08cm}
where  $\Pi^{f}$($\Pi^{b}$) is the spinon(holon) 
response function associated with the gauge field 
${\bf a}$(${\bf a}$ and ${\bf A}$);
$ \Pi^f_{X Y} = - \frac{ \partial^2 F^{f} }
{ \partial X \partial Y }$($ \Pi^b_{X Y} = - \frac{ \partial^2 F^{b} }
{ \partial X \partial Y }$);
the spinon(holon) response function associated with both 
the gauge fields and the spinon pairing field 
and $\Pi^0_{\Delta \Delta}$, the response function 
associated with the spinon pairing field.
It is shown that the first term 
represents the Ioffe-Larkin rule\cite{IOFFE} 
for the current response function
contributed only from  the gauge field fluctuations,
and the second term,  from the spin fluctuations.
Each contribution comes from the 
coupling between the charge and spin degrees of freedom,
as manifested by Eq.(\ref{Pi}).

The response function $\Pi^f$($\Pi^b$) is 
contributed from both the
paramagnetic and diamagnetic parts.
The paramagnetic response function is
obtained from the  current-current
correlation functions 
$ \Pi_{xx}^f(P)= <j^f_x(r^\prime ,t^\prime)
j^f_x(r,t )>-<j^f_x(r^\prime ,t^\prime)><j^f_x(r,t )>$ 
for the spinon and
$ \Pi_{xx}^b(P)= <j^b_x(r^\prime ,t^\prime)
j^b_x(r,t )>-<j^b_x(r^\prime ,t^\prime)><j^b_x(r,t )>$ 
for the holon,
and  the 
diamagnetic response function 
involves the average kinetic energy of spinon(holon).
$\Pi^f_{a \Delta}$($\Pi^b_{a \Delta}$) is given by the
correlations between the spinon(holon) current
and the 'anomalous' spinon(holon) pairing,
$\Pi^f_{a \Delta} = <j^f_x(r^\prime ,t^\prime) 
D^f (r,t )> - <j^f_x(r^\prime ,t^\prime)><D^f (r,t )>$
($\Pi^b_{a \Delta}=<j^b_x(r^\prime ,t^\prime) 
D^b (r,t )> - <j^b_x(r^\prime ,t^\prime)><D^b (r,t )>$)
with 
$D^f (r,t ) \sim \sum_l \left(
(e^{-i \tau} 
f_{r, \downarrow}(t)f_{r+l, \uparrow} (t) )
+ H.C. \right)$ and $D^b (r,t) \sim \sum_l
\left( b_r(t) b_{r+l}(t) + H.C.  \right) $.
 Here $l$ represents nearest neighbor sites
around location $r$ and 
$\tau = \pm \frac{\Pi}{2}$($+$(-) for x(y)-direction)
is a phase to represent the spinon pairing of d-wave symmetry. 
$\Pi_{\Delta \Delta}$ represents correlations between pairing currents;
$\Pi^f_{\Delta \Delta }=<D^f (r^\prime,t^\prime ) D^f (r,t ) >
-<D^f (r^\prime,t^\prime )><D^f (r,t )>$ 
for the spinon pairs and 
$\Pi^b_{\Delta \Delta}= <D^b (r^\prime,t^\prime ) D^b (r,t ) >
-<D^b (r^\prime,t^\prime )><D^b (r,t )>  $ for the holon pairs.

\begin{figure}
\centerline{\epsfig{file=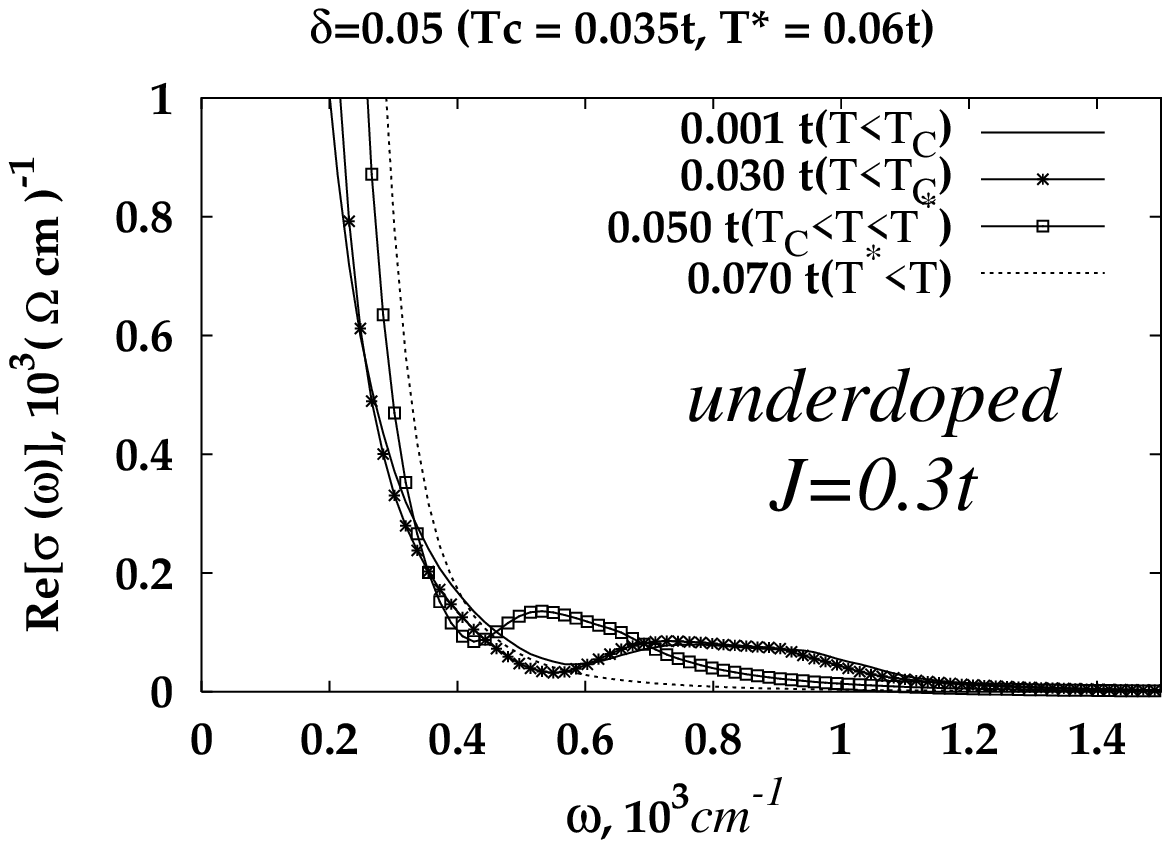,width=9cm,height=4.12cm, angle=0}}
\centerline{\epsfig{file=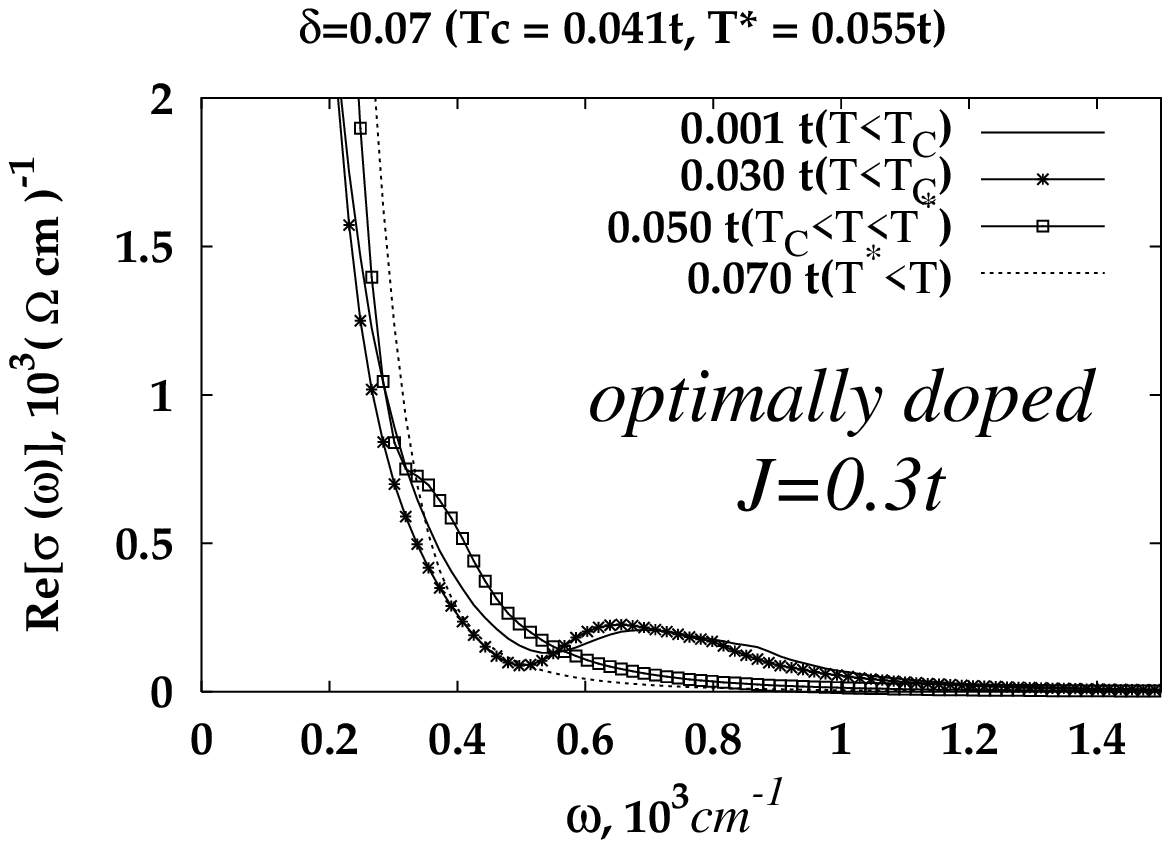,width=9cm,height=4.12cm, angle=0}}
\centerline{\epsfig{file=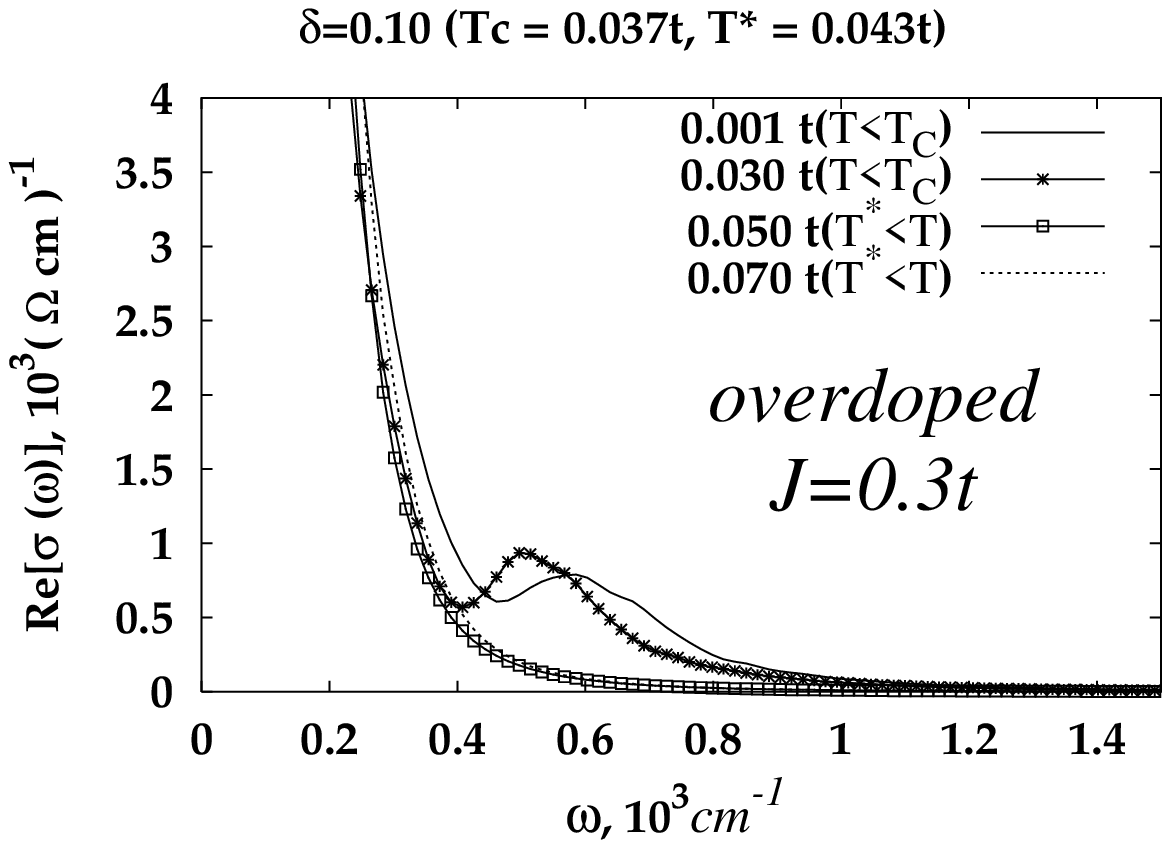,width=9cm,height=4.12cm, angle=0}}
\vspace{0.2cm}
\caption{}
\label{conductivity}
\end{figure}

\begin{figure}
\centerline{\epsfig{file=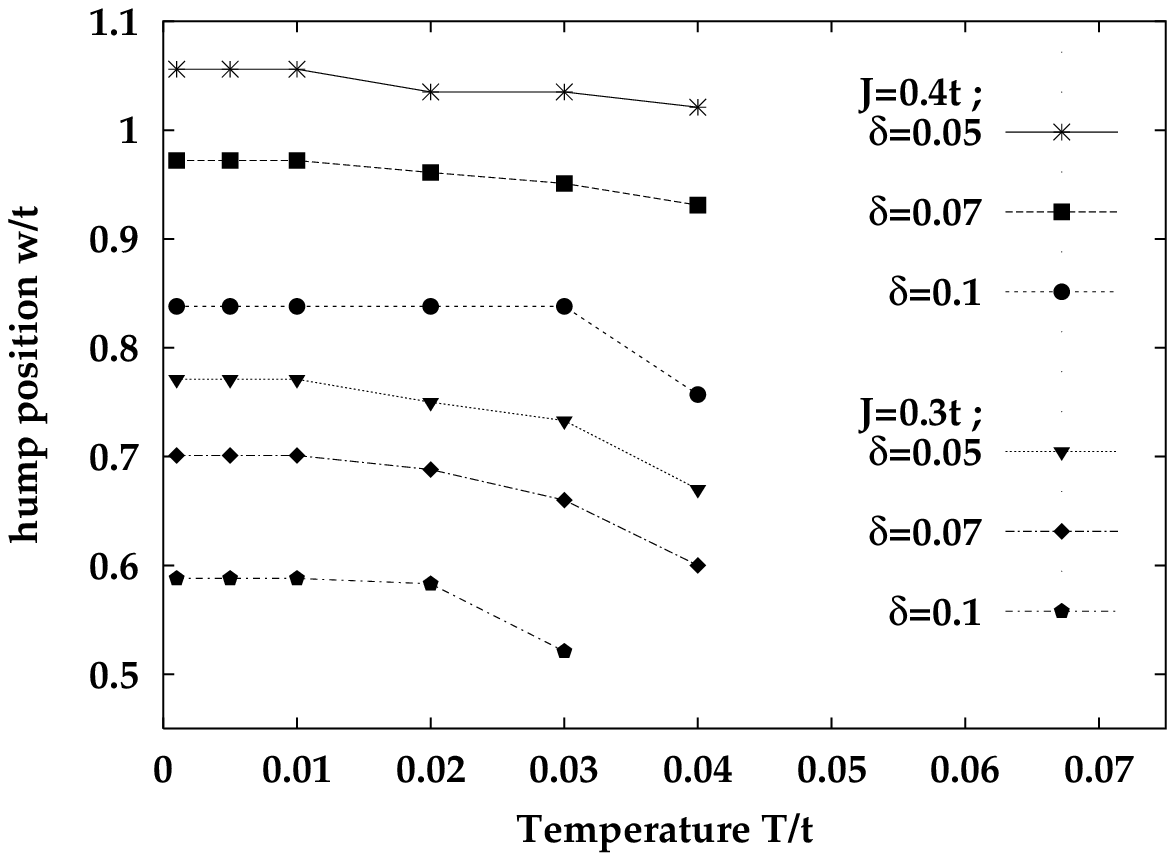, width=8cm, height=4.12cm,angle=0}}
\vspace{0.2cm}
\caption{}
\label{Hump}
\end{figure}

\begin{figure}
\centerline{\epsfig{file=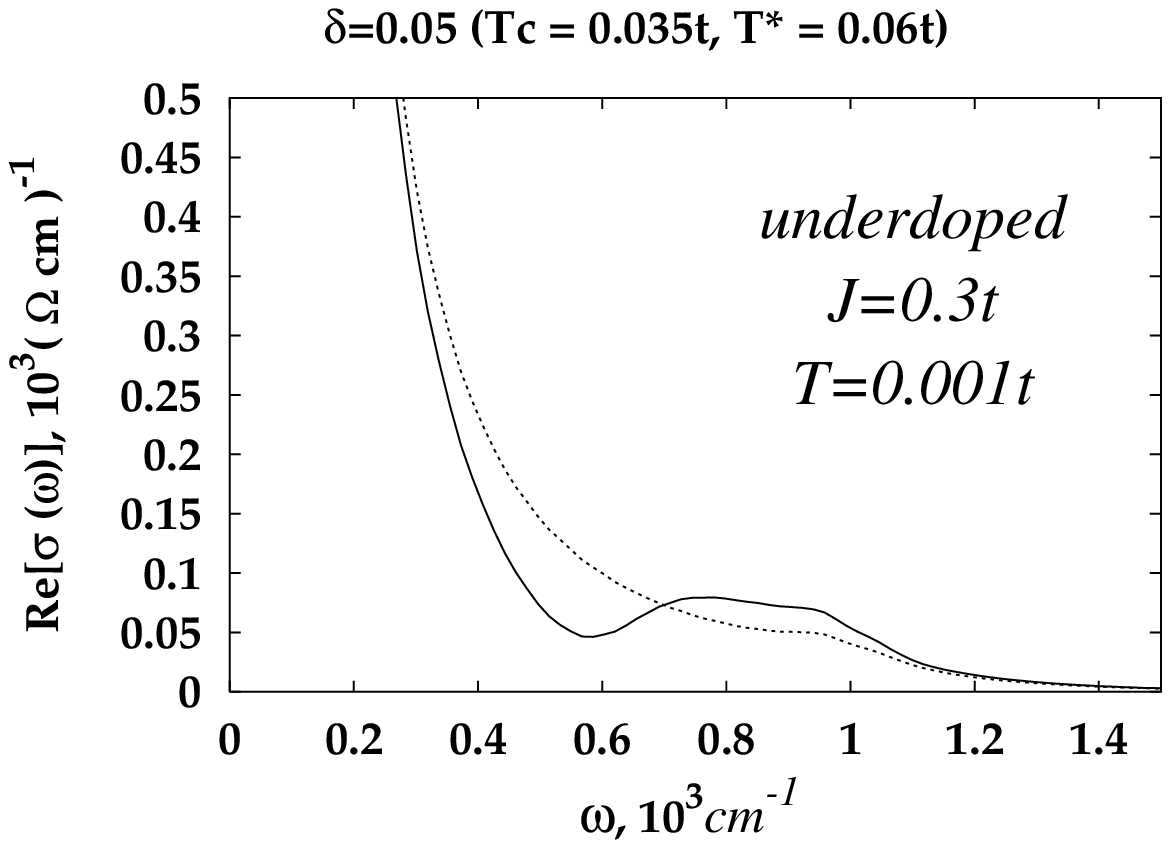, width=8cm, height=4.3cm,angle=0}}
\vspace{0.2cm}
\caption{}
\label{seperately}
\end{figure}

Fig.\ref{conductivity} 
shows 
 computed optical conductivities
from the U(1) slave-boson t-J Hamiltonian(Eq.(\ref{Hamiltonian_fb}))
with J=0.3t for the 
 underdoped($\delta=0.05$),
optimally doped($\delta=0.07$)
and overdoped($\delta=0.1$) regions.
Compared to the present U(1) result of optimal doping 
 the SU(2)
slave-boson theory\cite{SSLEE} predicted a more realistic 
value of optimal doping close to $\delta \approx  0.15$, by yielding a 
 phase diagram
of showing an  arch shaped bose condensation temperature 
in better agreement with observation.
To avoid complexity,  we resort to the simpler 
case of U(1) as our prime interest lies in 
the investigations of the
role of spin fluctuations and the 
coupling between the charge and spin degrees 
of freedom
on the formation of
peak-dip-hump structures, since the  accurate SU(2)
theory will not alter physics on the cause of 
the peak-dip-hump structure.
 Although not shown here for other values of J we find 
qualitative 
agreements with experiments
in that the 
 peak-dip-hump structures are well predicted
 below $T^*$ and $T_C$.
In Fig.\ref{Hump} the hump peak position is seen to remain nearly constant 
with the variation of hole doping and temperature 
below $T^*$
but not so above $T_C$.
In general, the predicted hump position 
tends to shift to a lower frequency with increasing hole concentration
and with temperature,
showing a gradual disappearance of the hump.
A trend of rapid drop in a high frequency region is seen
to be unrealistic.
In order to find the role of   spin fluctuations,
we neglected the second term in Eq.(\ref{Pi}).  
The hump structure(dotted line in Fig.\ref{seperately}) 
completely disappeared,
clearly  indicating that 
spin-spin correlations or spin fluctuations 
associated with the spin singlet excitations
are responsible for the hump formation
 in the optical conductivity(Fig.\ref{seperately}).
For an additional analysis of spin fluctuation 
 we computed the optical conductivity using 
the Lanczos exact diagonalization 
method for a two hole doped 
$4 \times 4$ lattice
by introducing various  Heisenberg antiferromagnetic 
coupling strength J. Despite the finite size effects 
an irregular but  gross feature of peak-dip-hump structure is
still  predicted indicating that the hump is
originated from the spin-spin correlations.
A linear increase in the hump position with J is 
predicted.
From both the slave-boson and Lanczos calculations 
we note 
that the peak locations of the hump 
are sensitive to the variation of the antiferromagnetic 
coupling strength J, by showing a linear increase. 
Further, as mentioned above 
the neglect of the spin fluctuations(the second term in Eq.(\ref{Pi}))
led to a sudden disappearance of the hump structure. 

Although not shown here, using the present U(1) slave-boson theory 
the predicted  spectral functions around 
$(\pi, 0 )$ point in momentum space also showed
the peak-dip-hump structure consistent with ARPES data.
This incoherent background or the hump around the $(\pi, 0 )$ point 
 was found to occur as a result of the antiferromagnetic spin fluctuations,
having a common feature with the hump structure of the optical conductivity.
Thus we conclude from these multifaceted studies that
the spin-spin correlations or the spin fluctuations
involved with electrons around the $(\pi, 0 )$ point
in momentum space are definitely the prime cause 
of the hump structures below $T^*$ and $T_C$.

In the present study, 
by paying attention to a wide range of both hole doping(under-,
optimal and over-doping) and temperature($T<T_C$, $T_C<T<T^*$,
and $T^*<T$) with no empirical parameters obtained
from measurements,
we examined the optical conductivity 
as a function of frequency
for the two-dimensional systems of strongly
correlated electrons. 
Allowing the coupling between  the spin and charge 
degrees of freedom as manifested in Eq.(\ref{Pi}),
the peak-dip-hump structures are
predicted in agreement with observations.
It is shown that the antiferromagnetic
spin fluctuations of short
range associated with the spin singlet pair excitations
are important in yielding the observed hump structure,
and that the hump position linearly increase with the
antiferromagnetic Heisenberg coupling strength.
In general, the  predicted peak-dip-hump structures are in good 
agreement with observations particularly in the temperature ranges of 
$T<T_C$ and $T_C<T<T^*$ 
for the underdoped case.
It is shown that 
  the spin fluctuations of the 
shortest possible antiferromagnetic correlation
length(that is the spin singlet pair) alone 
can cause the formation of the 
hump structure.
However, considerations of 
 both the antiferromagnetic
spin fluctuations of correlation lengths larger
than the spin singlet pair and the direct channel single spin 
fluctuations at high energies
may be needed
 to remedy quantitative discrepancies 
in the rapid drops of optical conductivity at 
temperatures above $T^*$
and at frequencies beyond the peak location of the hump.

One of us (SHSS) acknowledges the generous supports
of Korea Ministry of Education (Hakjin Excellence Leadership Program 2001)
and POSRIP Project at Pohang Unversity of Science and Technology

\newpage
\centerline{FIGURE CAPTIONS}
\begin{itemize}

\item[FIG. 1]
Computed optical conductivities 
as a function of temperature 
for $\delta = 0.05$(under doped), 
$\delta = 0.07$(optimally doped) and  $\delta = 0.1$(over doped) cases 
with the antiferromagnetic Heinsenberg coupling strength of J=0.3 for
all cases.

\item[FIG. 2]
Temperature dependence of hump position 
as a function of antiferromagnetic
coupling J and hole concentration.

\item[FIG. 3]
The total optical conductivity (solid line) vs. 
a partial one (dotted line) contributed only from the 
first term and thus from the neglect of the spin fluctuation (second) term in Eq.(\ref{Pi}).

\end{itemize}

\newpage

\begin{thebibliography}{22}
%
\bibitem{KOTLIAR} G. Kotliar and J. Liu, Phys. Rev. B {\bf 38}, 5142 (1988); references there-in.
%
\bibitem{FUKUYAMA} Y. Suzumura, Y. Hasegawa and H. Fukuyama, J. Phys. Soc. Jpn.
{\bf 57}, 2768 (1988)
\bibitem{UBBENS92} a) M. U. Ubbens and P. A. Lee, Phys. Rev. B {\bf 46}, 8434 (1992);
b) M. U. Ubbens and P. A. Lee, Phys. Rev. B {\bf 49}, 6853 (1994); references there-in.
%
\bibitem{WEN} a) X. G. Wen and P. A. Lee, Phys. Rev. Lett. {\bf 76}, 503 (1996);
b) X. G. Wen and P. A. Lee, Phys. Rev. Lett. {\bf 80}, 2193 (1998)
%
\bibitem{LEE} P. A. Lee, N. Nagaosa, T. K Ng and X. G. Wen,  Phys.  Rev. B,
{\bf 57}, 6003 (1998); 
N. Nagaosa  and Patrick A. Lee, Phys. Rev. B, {\bf 61}, 9166 (2000)
%
\bibitem{SSLEE}  S. -S. Lee and Sung-Ho Suck Salk, Phys. Rev. B {\bf 64} 052501 (2001);
Int. J. Mod. Phys. B {\bf 13},
3455 (1999); Physica C {\bf 353}, 130 (2001)
\bibitem{GIMM} T. H. Gimm, S. S. Lee, S. P. Hong
and Sung-Ho Suck Salk, Phys. Rev. B {\bf 60} 6324 (1999)
%
\bibitem{ROTTER} L. D. Rotter, Z. Schlesinger, R. T. Collins,
F. Holtzberg, and C. Field, Phys. Rev. Lett, {\bf 67}, 2741 (1991)
%
\bibitem{ROMERO} D. B. Romero, C. D. Porter, D. B. Tanner, L. Forro,
  D. Mandrus, L. Mihaly, G. L. Carr, and G. P. Williams, Phys. Rev. Lett. {\bf 68}, 1590 (1992) 
%
\bibitem{UCHIDA} S. Uchida, K. Tamasaku, K. Takenaka and Y. Fukuzumi,
J. Low. Temp. Phys. {\bf 105}, 723 (1996) 
%
\bibitem{PUCHKOV} A. V. Puchkov, D. N. Basov and T. Timusk,
J. Phys. Cond. Matt., {\bf 8}, 10049 (1996)
%
\bibitem{LIU} H. L. Liu, M. A. Quijada, A. M. Zibold, Y-D. Yoon, D. B. Tanner,
G. Cao, J. E. Crow, H. Berger, G. Margaritondo, L. Forro, Beom-Hoan O, 
J. T. Markert, R. J. Kelly and M. Onellion,
J. Phys. Cond. Matt., {\bf 11}, 239 (1999)
%
\bibitem{STOJKOVIC} Branko P. Stojkovi$\acute{c}$ and David Pines,
Phys. Rev. B, {\bf 56}, 11931 (1997); 
%
\bibitem{MUNZAR} D. Munzar, C. Bernhard and  M. Cardona,
Physica C, {\bf 312} 121 (1999)
%
\bibitem{HASLINER} R. Haslinger, Andrey V. Chubukov and Ar. Abanov,
Phys. Rev. B, {\bf 63}, 020503 (2000)
%
\bibitem{MBP} P. Monthoux, A. V. Balatsky, 
D. Pines, Phys. Rev. B, {\bf 46}, 14803 (1992); 
P. Monthoux, 
D. Pines, Phys. Rev. B, {\bf 47}, 6069 (1993); Phys. Rev. B, {\bf 49}, 4261 (1994)
%
\bibitem{CM} A. V. Chubukov and D. K. Morr, Phys. Rep, {\bf 288}, 355 (1997)
%

\bibitem{DAGOTTO} E. Dagotto, Rev. Mod. Phys. {\bf 66}, 763 (1994)
%
\bibitem{IOFFE} L. B. Ioffe, A. I. Larkin, Phys. Rev. B {\bf 39}, 8988 (1989)
%



\end{thebibliography}
\end{document}